# Engineered Kondo screening and nonzero Berry phase in SrTiO$_3$/LaTiO$_3$/SrTiO$_3$ heterostructures

Fang Yang[1,*], Zhenzhen Wang[1,2], Yonghe Liu[1,2], Shuai Yang[1], Ze Yu[1,2], Qichang An[1], Zhaoqing Ding[1,2], Fanqi Meng[1], Yanwei Cao[3], Qinghua Zhang[1], Lin Gu[4], Miao Liu[1], Yongqing Li[1], Jiandong Guo[1,2,5,*], Xiaoran Liu[1,*]

1. Beijing National Laboratory for Condensed Matter Physics and Institute of Physics, Chinese Academy of Sciences, Beijing 100190, P. R. China
2. School of Physical Sciences, University of Chinese Academy of Sciences, Beijing 100049, China
3. Ningbo Institute of Materials Technology and Engineering, Chinese Academy of Sciences, Ningbo 315201, China
4. Beijing National Center for Electron Microscopy and Laboratory of Advanced Materials, Department of Materials Science and Engineering, Tsinghua University, Beijing 100084, China
5. Songshan Lake Materials Laboratory, Dongguan 523808, China

**Abstract**

Controlling the interplay between localized spins and itinerant electrons at the oxide interfaces can lead to exotic magnetic states. Here we devise SrTiO$_3$/LaTiO$_3$/SrTiO$_3$ heterostructures with varied thickness of the LaTiO$_3$ layer ($n$ monolayers) to investigate the magnetic interactions in the two-dimensional electron gas system. The heterostructures exhibit significant Kondo effect when the LaTiO$_3$ layer is rather thin ($n$ = 2, 10), manifesting the strong interaction between the itinerant electrons and the localized magnetic moments at the interfaces, while the Kondo effect is greatly inhibited when $n$ = 20. Notably, distinct Shubnikov-de Haas oscillations are observed and a nonzero Berry phase of $\pi$ is extracted when the LaTiO$_3$ layer is rather thin ($n$ = 2, 10), which is absent in the heterostructure with thicker LaTiO$_3$ layer ($n$ = 20). The observed phenomena are consistently interpreted as a result of sub-band splitting and symmetry breaking due to the interplay between the interfacial Rashba spin-orbit coupling and the magnetic orderings in the heterostructures. Our findings provide a route for exploring and manipulating nontrivial electronic band structures at complex oxide interfaces.

# Introduction

The complex oxide interface has been revealed as a remarkable playground for the exploration of intriguing collective electronic and magnetic phenomena [1-5]. In addition to the lattice, charge, orbital, and spin degrees of freedom inherited from constituent materials, unique opportunities can be provided at the coherently formed interfaces, triggering emergent or hidden properties inaccessible in bulk compounds [6,7]. In particular, at an oxide interface with intrinsic inversion symmetry breaking, the Rashba spin-orbit coupling (SOC) is naturally induced due to the potential drop across the interface, lifting the spin degeneracy of the $d$ electrons and locking the spins to the linear momentum. Such a "Rashba spin splitting" scenario has been recognized as playing key roles in a vast variety of developing fields in spintronics, orbitronics, and topological quantum states of matter [8].

As one of the most representative phenomena, the two-dimensional electron gases (2DEGs) at complex oxide interfaces have attracted tremendous attention over the past 20 years [9-12]. Unlike the situations in conventional II-VI or III-V semiconductor quantum wells with $s$ or $p$ mobile electrons, 2DEGs with the correlated $d$ electrons have exhibited fascinating emergent features including superconductivity [13,14], ferromagnetism [15,16], Kondo effect [17], significant charge transfer [18], spin and orbital polarizations [19]. Recently, more interest has been focused on investigating and manipulating the effect of the Rashba SOC in these systems [20-25]. Specifically, Caviglia *et al.* reported the discovery of a large Rashba SOC at the interface of LaAlO$_3$/SrTiO$_3$, whose magnitude can be effectively modulated by an external gate voltage [20]. Herranz *et al.* revealed that the Rashba SOC and the corresponding band splitting at LaAlO$_3$/SrTiO$_3$ interface is tunable via the selection of the crystal orientations [21]. Lin *et al.* fabricated asymmetric LaAlO$_3$/SrTiO$_3$/LaAlO$_3$ quantum wells with opposite polar discontinuities at the top and bottom interfaces, and found a transition from the cubic Rashba effect to the coexistence of linear and cubic Rashba effects [24]. Very recently, Li *et al.* studied systematically the effect of Rashba SOC at $R$AlO$_3$/SrTiO$_3$ ($R$ = La, Pr, Nd, Sm, and Gd) interfaces [25].

Compared to LaAlO$_3$, LaTiO$_3$ exhibits more complicated features from several aspects. First, LaTiO$_3$ is a Mott insulator (Ti$^{3+}$, 3$d^1$) with a small gap of only ~0.1 eV [26]. Second, it undergoes a G-type antiferromagnetic (AFM) phase transition below ~146 K [27-29]. Third, unlike the termination-dependent doping of SrTiO$_3$ ($p$- or $n$-type) at the LaAlO$_3$/SrTiO$_3$ interface, the 3$d$ electrons are intrinsically transferred from LaTiO$_3$ to SrTiO$_3$ at their interfaces, leading to remarkably enhanced amounts of localized magnetic moments [30]. Thereby, the LaTiO$_3$-based 2DEG system is an ideal platform to understand the role of interfacial Rashba SOC in the magnetic interactions, as well as how it affects the topology of the low-dimensional electronic band structures.

In this work, we synthesized a set of (001)-oriented SrTiO$_3$/LaTiO$_3$/SrTiO$_3$ heterostructures with the thickness of LaTiO$_3$ layers ($n$ monolayers) varying from 2 to 20 monolayer (ML) and investigated their electronic and magnetic properties via various transport measurements. Distinct Kondo screening of the localized moments by the mobile carriers is observed in samples with ultrathin LaTiO$_3$ layer ($n$ = 2 and 10), whereas the effect is much suppressed when the LaTiO$_3$ layer is sufficiently thick ($n$ = 20). The latter indicates the formation of the bulk-like AFM ordering by which the localized moments are pinned or polarized, resulting in the suppression of scattering. More intriguingly, samples with ultrathin LaTiO$_3$ layer exhibit a nonzero Berry phase of π of the conductive electrons as deduced from the Shubnikov-de Haas (SdH) oscillations, which becomes zero in thick one. These phenomena can be consistently interpreted as a result of the interplay of the Rashba spin splitting scenario plus the breaking of the time-reversal symmetry in the heterostructures. These findings open new pathways for engineering the Rashba SOC and manipulating the nontrivial band topology at low-dimensional systems.

**Results**

The SrTiO$_3$/LaTiO$_3$/SrTiO$_3$ heterostructures [see Fig. 1(a)] were fabricated on SrTiO$_3$ (001) substrates by pulsed laser deposition equipped with *in-situ* reflection high energy electron diffraction (RHEED). The thickness of the LaTiO$_3$ layers $n$ was varied from 2 to 20 MLs, while the thickness of the SrTiO$_3$ quantum wells on both sides are fixed at 10 MLs. Before the deposition, TiO$_2$-terminated SrTiO$_3$ (001) substrates were prepared by etching in buffered hydrofluoric acid for 45 s, followed by annealing in an oxygen atmosphere at 950 °C for 2 hours. The RHEED pattern of the treated SrTiO$_3$ substrate is displayed in Fig. 1(b) inset (left panel). To prevent from oxidizing LaTiO$_3$ to La$_2$Ti$_2$O$_7$ or any other oxygen-rich phases, we grew the LaTiO$_3$ layers under a high vacuum condition with the pressure of ~10$^{-7}$ mbar at a substrate temperature of ~720°C measured by a pyrometer. The heterostructures were established in the layer-by-layer growth mode, as evidenced by the distinct oscillations of RHEED intensity shown in Fig. 1b. This allows us to precisely control the thickness of each compound in the heterostructures by counting the number of oscillations during the deposition process. The RHEED pattern after the deposition [see right panel of the inset in Fig. 1(b)], which exhibits practically the same pattern as the substrate, indicates the establishment of flat and well-crystallized surfaces of the films. The morphology of the heterostructures is further revealed by atomic force microscopy imaging [Fig. 1(c)], which demonstrates the formation of high-quality films with well-defined steps and terraces on the surface.

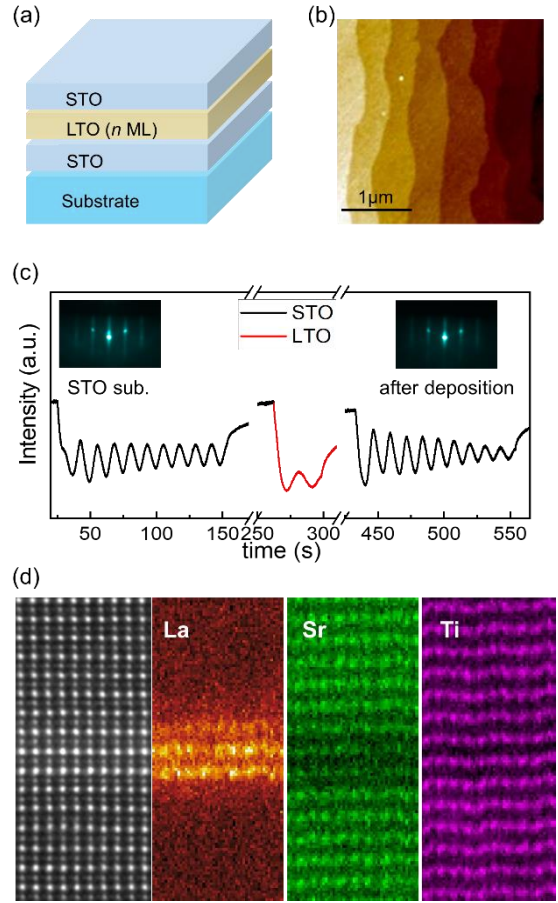

Figure 1 (a) Schematic of the SrTiO$_3$(10 ML)/LaTiO$_3$($n$ ML)/SrTiO$_3$(10 ML) heterostructures. (b) RHEED oscillations of SrTiO$_3$(10 ML)/LaTiO$_3$(2 ML)/SrTiO$_3$(10 ML) during growth. The black and red curves are the oscillations of SrTiO$_3$ and LaTiO$_3$ during growth, respectively. Insets: the RHEED patterns of the treated SrTiO$_3$ substrate and the topmost as-grown SrTiO$_3$ film, respectively. (c) The morphology of the topmost as-grown SrTiO$_3$ films measured by Atomic Force Microscopy. (d) HAADF-STEM image along [110] direction and the corresponding EDX mappings of all cations (La, Sr and Ti) across the interfaces, showing the distinguishable La distributions of SrTiO$_3$/LaTiO$_3$ (2 ML)/SrTiO$_3$.

To investigate the interfaces of the heterostructures, cross-sectional scanning transmission electron microscopy (STEM) imaging was performed in the high-angle annular dark-field (HAADF) mode (see Fig. S1 in the Supplemental Material [31]). A representative image of the SrTiO$_3$/LaTiO$_3$ (2 ML)/SrTiO$_3$ heterostructure is shown in Fig. 1(d). The large (white) spots represent the Sr and La columns, while the small (gray) ones for the Ti columns. Notably, the heterostructure is demonstrated with high crystallinity, atomically sharp interfaces, and uniformly distributed LaTiO$_3$ with precisely two monolayers among the SrTiO$_3$ quantum wells. In addition, the associated energy-dispersive x-ray spectroscopy (EDX) mappings of all cations (La, Sr and Ti) across the interfaces indicate the interdiffusion of La and Sr atoms is limited within 1 ML at the interfaces.

After demonstrating the structural quality, we turn to examine their electrical properties. The temperature dependence of the resistance was measured using the four-point probe method on each sample ($n$ = 2, 10, 20) with the application of an external magnetic field along normal direction of the surface. While a typical parabolic R vs. T behavior of normal metals is observed on each sample at relatively high temperatures, we pay special attention to the behavior at low temperatures (30 - 0.3 K). As seen from the normalized sheet resistance R(T)/R(0.3 K) in Fig. 2 (a)-(c), a remarkable feature is the pronounced upturn and saturation of the sheet resistance at the lower temperature after a resistance minimum, characteristic of the Kondo effect [15, 34-37]. Indeed, each curve can be well described by the Kondo fitting, regardless of the LaTiO$_3$ thickness. Strikingly, as the magnetic field increases, the behavior of the resistance upturn and saturation of each sample becomes more prominent. Such behaviors may originate from the enhanced forward scattering of diffusive electrons by the spin-orbit interaction [36].

It is noteworthy that while other factors such as weak localization [38] or electron-electron interactions [39] may also give rise to an increment of resistance at low temperatures, the Kondo effect is particularly distinguished in that it satisfies the universal scaling behavior [40]. Namely, with the numerical renormalization group (NRG) theory, the electrical resistance induced by the Kondo effect can be expressed by a universal function with only a single variable $T/T_K$ [40-42]:

$$R(T) = R_0 + \alpha T^2 + \beta T^5 + R_K(T/T_K)$$

$$R_K\left(\frac{T}{T_K}\right) = R_{K0}\left(\frac{1}{1+(2^{\frac{1}{s}}-1)(T/T_K)^2}\right)^s$$

Here $T_K$ is the Kondo temperature; $R_0$, $\alpha T^2$ and $\beta T^5$ with $\alpha$, $\beta$ being constants represent the contributions from residual resistance, the electron-electron interaction, and the electron-phonon interaction, respectively; $R_K(T/T_K)$ is an empirical function for the universal Kondo resistance with respect to the variable $T/T_K$; $R_{K0}$ is the Kondo resistance at zero temperature and the parameter $s$ is the effective spin of the magnetic scattering centers. (Here we take the same $s = 0.75$ as that taken in other titanate 2DEG systems [22]).

As displayed in Figure 2(d), all of the renormalized results fall onto a single theoretical curve, obeying the universal Kondo scaling very well. These results thus demonstrate the Kondo effect as the dominant factor for the upturn and saturation behavior of R(T) at low temperatures, where the mobile carriers experience the Kondo-like scattering by the localized magnetic moments at both interfaces formed by LaTiO$_3$ and SrTiO$_3$. The derived Kondo temperature $T_K$ of each sample is summarized in Fig. 2(e). For the $n$ = 10 and 2 heterostructures, $T_K$ decreases monotonically with the increase of the external magnetic field. In contrast, however, $T_K$ increases slightly with the increase of external magnetic field in $n$ = 20.

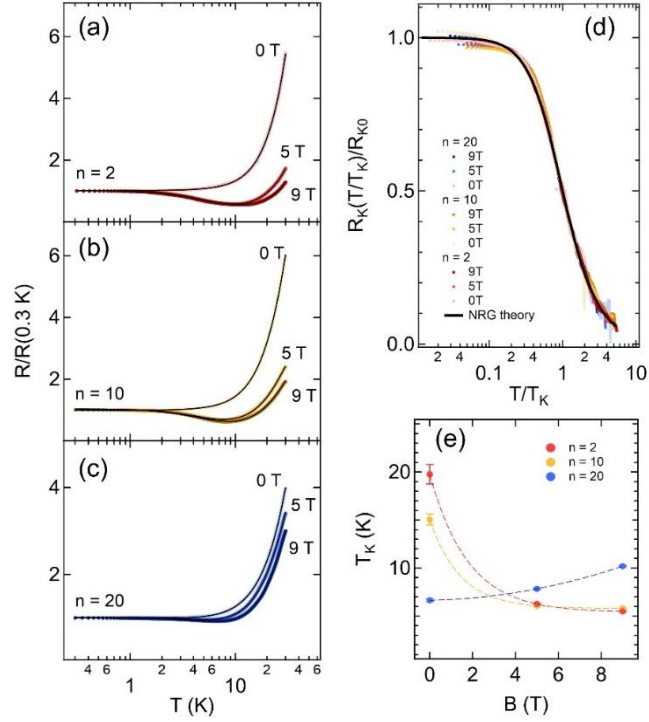

Figure 2 Normalized resistances R(T)/R(0.3 K) of SrTiO$_3$/LaTiO$_3$($n$ ML)/SrTiO$_3$ samples with (a) $n$ = 2, (b) $n$ = 10, (c) $n$ = 20 under different magnetic fields and the corresponding fitting results (black lines). (d) Experimentally and theoretically [black line, numerical renormalization group (NRG)] scaled Kondo resistances [R$_K$/(T/T$_K$)/R$_{K0}$]. (e) Kondo temperature T$_K$ as a function of magnetic field in each heterostructure with different $n$. The dashed lines are guide to the eye.

After investigating the temperature-dependent properties, we focused on exploring the ground state of each sample by measuring the field-dependent resistance at the base temperature 0.3 K. The magnetic field was applied perpendicular to the film surface during the experiments. As displayed in Fig. 3(a), an overall positive magnetoresistance (resistance increases as field increases) is observed on all samples, which shows no sign of saturation up to 9 T. Remarkably, compared to the weak linear magnetoresistance (MR) behavior in $n$ = 20, it is much more pronounced in $n$ = 10 and 2 with clear signatures of oscillations at large fields. Figure 3 (b)-(c) exhibit the distinct Shubnikov-de Haas (SdH) oscillations in $n$ = 10 and 2 heterostructures as a function of the inverse magnetic field 1/B after subtracting the background from Fig. 3(a). Note, no SdH oscillations are observed in $n$ = 20. The corresponding frequencies of these quantum oscillations in $n$ = 10 and 2 heterostructures are 23.8T and 26.3T, respectively.

Besides as a fingerprint of 2DEGs with high carrier mobility, measurements of the SdH oscillations can also reveal information about the topology of the electronic band structures and provide insights into the Berry phase of the mobile carriers [43]. According to the Lifshitz-Onsager quantization rule, the Berry phase can be derived from the Landau level fan diagram by extrapolating the dependence of 1/$B$ on the Landau index to 1/$B$ = 0. The Lifshitz-Onsager quantization rule is written as follows: $A\frac{\hbar}{eB} = 2\pi(N + \frac{1}{2} - \frac{\varphi}{2\pi})$, where A is the extremal cross-sectional area of the Fermi surface perpendicular to the applied magnetic field, $B$ is the strength of the magnetic field, $N$ is the Landau level index, and $\varphi$ is the Berry phase of the orbit in $k$-space. The obtained Landau fan diagrams for $n = 10$ and 2 heterostructures are shown in Figures 3(d) and 3(e), respectively. Note, the integer indices refer to the $R_{xx}$ peak positions with respect to $1/B$, while the half-integer indices to the $R_{xx}$ valley positions. The intercept of the best linear fit of $1/B$ versus $N$ for both heterostructures is practically zero, leading to a characteristic nonzero Berry phase of $\pi$ in both samples (~$1.05\pi$ for $n = 10$ and ~$1.01\pi$ for $n = 2$).

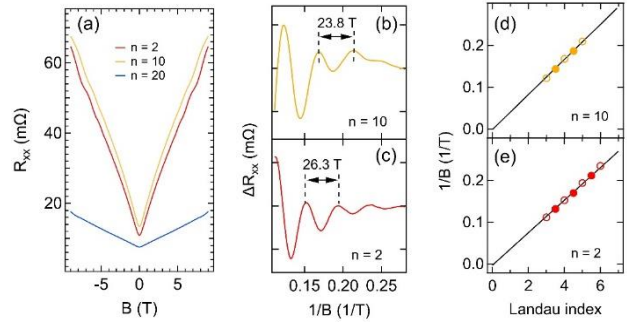

Figure 3 (a) Resistance of SrTiO$_3$/LaTiO$_3$/SrTiO$_3$ with various $n$ as a function of magnetic field at 0.3 K. (b) – (c) The SdH oscillations derived from MR in SrTiO$_3$/LaTiO$_3$/SrTiO$_3$ for (b) $n = 10$; (c) $n = 2$. The corresponding frequencies of the oscillations are 23.8 T ($n = 10$) and 26.3 T ($n = 2$), respectively. (d) – (e) Landau level fan diagrams of (d) $n = 10$ and (e) $n = 2$ heterostructures. Maxima and minima in $\Delta R_{xx}$ are denoted by open and solid circles, respectively.

**Discussion**

To exclude the possible conductivity contributions from LaTiO$_3$ or SrTiO$_3$ bulks, we note that the temperature dependence of resistivity of a homoepitaxial SrTiO$_3$ film ($n = 0$) exhibits a semiconducting behavior, whereas all the LaTiO$_3$/SrTiO$_3$ heterostructures exhibit metallic behaviors (Fig. S2 [31]). In addition, we prepared single LaTiO$_3$ films under the same growth condition (720°C, $10^{-7}$ mbar) on LaAlO$_3$ and (LaAlO$_3$)$_{0.3}$(Sr$_2$AlTaO$_6$)$_{0.7}$ (LSAT) substrates, respectively. Both samples are insulating (the room-temperature resistance is larger than 200MΩ). The formation of La$_2$Ti$_2$O$_7$ phase can all be safely ruled out by X-ray diffraction results. These observations are consistent with previous results [44], indicating the absence of the strain-induced conductivity inside LaTiO$_3$, as similarly mentioned in Ref. 45.

In addition, the spatial distribution of Ti$^{3+}$ across the interfaces in $n = 2$ sample has been extracted from the electron energy loss spectroscopy (EELS) of STEM (Fig. S3 [31]). The fraction of Ti$^{3+}$ [Ti$^{3+}$/(Ti$^{3+}$+Ti$^{4+}$)] reaches a maximal value of ~42% in the LaTiO$_3$ layers and gradually decays as entering into the SrTiO$_3$ layers on both sides. This behavior agrees well with the scenario of interfacial charge transfer, and the values are consistent with the results in literatures [9]. Thus, we can confirm that the origin of the conductivity is from the interface, as a result of the spontaneous electronic reconstruction where the $d$ electrons of LaTiO$_3$ are transferred to fill the empty $t_{2g}$ orbitals of SrTiO$_3$. [9, 16, 46]. According to the EELS results, the ratio of Ti$^{3+}$/(Ti$^{3+}$+Ti$^{4+}$) for $n = 2$ sample is about 42% such that the average valence of Ti is +3.6. This means ~0.6 electron per Ti site of LaTiO$_3$ is transferred to SrTiO$_3$. This value is consistent with previous reports [9, 16, 44]. A sheet carrier density $n_{EELS} \approx 10^{14}$/cm$^2$ is derived from the EELS results. On the other hand, the carrier density estimated from SdH oscillations is $n_{SdH} \approx 10^{12}$/cm$^2$. Such a discrepancy is likely due to two factors: (1) Some of the transferred electrons become localized on the SrTiO$_3$ side; (2) As reported by Veit et al. [23], the quantum oscillations originate from the high-mobility carriers from the hybridized $d_{xz/yz}$ bands, while the low-mobility carriers from the $d_{xy}$ bands are not captured.

Next, it is noteworthy that unlike in SrTiO$_3$/LaAlO$_3$/SrTiO$_3$ systems where the asymmetric sequences of stacking layers "SrO-AlO$_2$-LaO" ("SrO-TiO$_2$-LaO") give rise to "p-type" ("n-type") interface [24], the interfaces in SrTiO$_3$/LaTiO$_3$/SrTiO$_3$ are identical. This can be further revealed from the EELS results (Fig. S3 [31]), where the charge transfer occurs at both interfaces with a symmetric distribution of Ti valency.

The detailed band structures induced by Rashba splitting can be rather complicated at oxide interfaces [47-49]; nevertheless, the observed phenomena can be rationalized by a microscopic model in which the electronic band topology is determined by the interplay between the interfacial Rashba SOC and magnetic interactions. Specifically, the charge transfer process at the LaTiO$_3$/SrTiO$_3$ interface leads to the coexistence of both mobile electrons and uncompensated localized moments on Ti sites at the interfacial regions. [36, 50] [see Fig. 4(a)]. Consequently, the screening of these localized magnetic moments by the conducting carriers takes place at low temperatures, which accounts for the observed Kondo effect in these heterostructures. On the other hand, it is notable that the inversion asymmetry at the heterointerfaces would naturally trigger the scenario of the Rashba spin splitting [23]. As a result, the doubly degenerate Ti $t_{2g}$ energy bands are split, leading to the formation of a band crossing at the Γ point [left panel of Fig. 4(b)]. Such a band crossing with opposite spin characters can cause a strong Berry curvature in the momentum space, and for $n$ = 10 and 2 the mobile carriers gain a nonzero Berry phase [left panel of Fig. 4(c)].

When it comes to the heterostructure with $n$ = 20, both Kondo effect and magnetoresistance exhibit different behaviors from those in $n$ = 2 and 10. Note, previous density functional theory calculations on a superlattice composed of (SrTiO$_3$)$_4$/(LaTiO$_3$)$_6$ suggested the recovery of AFM ordering in 6 ML LaTiO$_3$ slab [51]. However, considering the depth of charge transfer extending up to 4-5 ML into LaTiO$_3$ at each interface [9] combined with the enhanced quantum fluctuations at low dimensions, the AFM ordering is probably not recovered in $n$ = 10 sample. Nevertheless, it is reasonable to expect that in $n$ = 20, the LaTiO$_3$ slab is thick enough to establish the bulk-like antiferromagnetism. [51] [right panel of Fig. 4(a)].

The long-range AFM ordering in $n$ = 20 LaTiO$_3$ not only breaks the time-reversal symmetry of the system, but also changes the internal molecular field which further affects the localized magnetic moment near the interfaces. In particular, the randomly oriented spins which would generate strong scattering are greatly suppressed when the moments are pinned or polarized by the molecular field, accounting for the observed fading of the Kondo characteristics [52]. As can be seen in Fig. 2(c), the trend of the Kondo effect in $n$ = 20 is not as prominent as that in $n$ = 10 and 2. Once a magnetic field is applied, it may help to counteract the effect of the AFM ordering in the heterostructures with thick LaTiO$_3$ layers ($n$ = 20); whereas it conversely polarizes the localized magnetic moments in the heterostructures with $n$ = 10 and 2 and weakens the magnetic scattering. In this manner, the field dependence of $T_K$ is quite different between the heterostructures with thin and thick LaTiO$_3$ layers; i.e., $T_K$ increases with the increase of magnetic field for $n$ = 20, while it decreases with magnetic field for both $n$ = 10 and 2 heterostructures. In the heterostructure with $n$ = 20, the interfacial Rashba SOC still lifts the double degeneracy of the hybridized $d_{xz/yz}$ band, similar to the case in heterostructures with $n$ = 2 and 10. Nevertheless, breaking of the time-reversal symmetry by the long-range AFM ordering renders the Kramers degeneracy is further split, with the formation an energy gap at the Γ point [right panel of Fig. 4(b)]. In this manner, the Berry phase returns zero in $n$ = 20 [right panel of Fig. 4(c)]. We note that although the carrier densities may slightly vary in different samples, it affects the strength of SOC and results in the variation of the splitting between those two crossing Rashba sub-bands [20]. However, the crossing of these two sub-bands at the Γ point is still robust in the momentum space and cannot be annihilated.

In the end, we would like to emphasize the distinction of our LaTiO$_3$/SrTiO$_3$/LaTiO$_3$ quantum wells compared to the bilayer LaTiO$_3$/SrTiO$_3$ heterostructure (LaTiO$_3$ layer interfaces with vacuum) by M. Veit *et al.* [23]. Although a nonzero Berry phase of π has been observed in both structures which reconciles qualitatively with the general interfacial Rashba SOC scenario, the quantum wells geometry lends a significant impact on the electronic structures of the 2DEGs. On one hand, the quantum wells geometry provides an extended pathway for electrons hopping via the Ti-O-Ti bonds, suppressing the formation of localized moments and long-range magnetic orderings. This can explain why there were no more SdH oscillations in LaTiO$_3$/SrTiO$_3$ with LaTiO$_3$ thicker than 5 ML [23], whereas in the present work the oscillations can be preserved even in heterostructures with 10 ML-thick LaTiO$_3$. On the other hand, the position of the Rashba split hybridized $d_{xz/yz}$ bands relative to the Fermi level is plausibly varied. The distortions of the TiO$_6$ octahedra at both interfaces of the quantum wells might slightly lift the $d_{xz/yz}$ bands, such that the Fermi level only intersects the outer Rashba split band. As a result, a single frequency is observed in the quantum wells, whereas two sets of frequencies were observed in bilayers due to the intersection of the Fermi level with both inner and outer Rashba split bands [23].

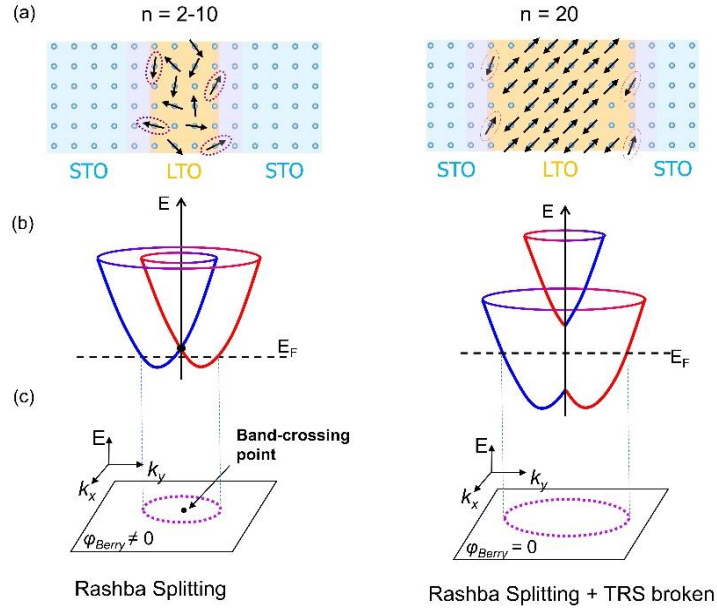

Figure 4 Schematic illustration of the electronic band structures determined by the interplay between magnetic interaction and interfacial Rashba SOC in SrTiO$_3$/LaTiO$_3$($n$ ML)/SrTiO$_3$ heterostructures. (a) Schematic illustration of the localized magnetic moments at the 2DEG interfaces and their screening by mobile carriers. The blue balls refer to the Ti atoms. The pink shadow regions represent the 2DEG with itinerant electrons, and the black arrows

and those circled by purple dotted ellipses refer to the magnetic moments in LaTiO$_3$ layer and the localized moments at the interfaces, respectively. Left panel: for the ultrathin cases ($n$ = 2-10) without AFM order. Right panel: for the thick case ($n$ = 20) with AFM ordering breaking time-reversal symmetry. (b) The electronic band structures in the heterostructures. The purple dashed ellipses represent the orbits of the carriers. Left panel: the splitting of the doubly degenerate Ti $t_{2g}$ energy bands leads to a band crossing at the Γ point. Right panel: with the breaking of the time-reversal symmetry by the long-range AFM ordering, a gap opens at the Γ point. (c) The projection of (b). Left panel: the contour encloses the band-crossing point, leading to a $π$ Berry phase. Right panel: the band-crossing point is annihilated by time reversal symmetry broken, and the Berry phase is zero.

## Summary

To summarize, we devise a set of (001)-oriented SrTiO$_3$/LaTiO$_3$ ($n$ ML)/SrTiO$_3$ heterostructures with varied thicknesses of the LaTiO$_3$ layer. Two-dimensional electron gases are formed at both interfaces, where the carriers strongly interact with the localized magnetic moment, leading to a significant Kondo effect. Distinct Shubnikov-de Haas oscillations are observed and a nonzero Berry phase of $π$ is extracted when the LaTiO$_3$ layer is rather thin ($n$ = 2, 10), which is absent for the thicker LaTiO$_3$ layer ($n$ = 20). The observed phenomena are consistently interpreted as a result of the symmetry breaking due to the interplay between the interfacial Rashba spin-orbit coupling and the magnetic orderings in the heterostructures. Our findings thus provide a novel avenue for exploring and manipulating nontrivial electronic band structures at complex oxide interfaces.

## Acknowledgements

We thank the staff from BL07U beamline of Shanghai Synchrotron Radiation Facility (SSRF) for assistance of x-ray absorption spectroscopy data collection. This work was supported by the National Key R&D Program of China (No. 2017YFA0303600), the National Natural Science Foundation of China (No. 11974409, 11874058, 52072400, 52025025), and the Strategic Priority Research Program (B) of the Chinese Academy of Sciences (No. XDB33000000).